# Review of the August 1972 and March 1989 Space Weather Events: Can We Learn Anything New From Them?

Bruce T. Tsurutani[1*], Abhijit Sen[2], Rajkumar Hajra[3], Gurbax S. Lakhina[4], Richard B. Horne[5], and Tohru Hada[6]

[1]Retired, Pasadena, California USA.

[2]Institute for Plasma Research, Gandhinagar, India.

[3]CAS Key Laboratory of Geospace Environment, School of Earth and Space Sciences, University of Science and Technology of China, Hefei, China.

[4]Retired, Vashi, Navi Mumbai 400703, India.

[5]British Antarctic Survey, Cambridge, UK.

[6]International Research Center for Space and Planetary Environmental Science, Kyushu University, Fukuoka, Japan.

[*]Corresponding author: Bruce Tsurutani (bruce.tsurutani@gmail.com)

**Key Points:**

- The August 1972 and March 1989 space weather events are summarized and compared to the Carrington plus a few other recent events.

- With different solar and interplanetary conditions, the August 1972 ICME could have caused a storm greater than the Carrington storm.

- The complex March 1989 storm was a "stealth magnetic storm", probably greater ring-current energy-wise than the Carrington storm.

## Abstract

Updated summaries of the August 1972 and March 1989 space weather events have been constructed. The features of these two events are compared to the Carrington 1859 event and a few other major space weather events. It is concluded that solar active regions release energy in a variety of forms (X-rays, EUV photons, visible light, coronal mass ejection (CME) plasmas and fields) and they in turn can produce other energetic effects (solar energetic particles (SEPs), magnetic storms) in a variety of ways. It is clear that there is no strong one-to-one relationship between these various energy sinks. The energy is often distributed differently from one space weather event to the next. Concerning SEPs accelerated at interplanetary CME (ICME) shocks, it is concluded that the Fermi mechanism associated with quasi-parallel shocks is relatively weak and that the gradient drift mechanism (electric fields) at quasi-perpendicular shocks will produce harder spectra and higher fluxes. If the 4 Augusut 1972 intrinsic magnetic cloud condition (southward interplanetary magnetic field instead of northward) and the interplanetary Sun to 1 au conditions were different, a 4 August 1972 magnetic storm and magnetospheric dawn-to-dusk electric fields substantially larger than the Carrington event would have occurred. Under these special interplanetary conditions, a Miyake et al. (2012)-like extreme SEP event may have been formed. The long duration complex 1989 storm was probably greater than the Carrington storm in the sense that the total ring current particle energy was larger.

## Plain Language Summary

All the main information for the August 1972 and March 1989 space weather events have been gathered together for the readership. Various features of these two events have been compared and contrasted to the Carrington event and several other recent major events.

It is thought that major energetic MeV to GeV particle acceleration occurs at interplanetary shocks instead of acceleration at the solar flare site. Two popular mechanisms for shock particle acceleration are: 1) Fermi acceleration between waves upstream and downstream of a shock, and 2) particle gradient drift in the direction of an electric field along the shock surface. The second mechanism is concluded to form a harder particle spectrum with higher fluxes.

If the intrinsic conditions of the 4 August 1972 coronal mass ejection were different (southward interplanetary magnetic fields) and that propagated to the Earth without major evolution or distortion, a magnetic storm equal to or more intense than the Carrington event would have occurred. We have also constructed a scenario involving altered interplanetary space for the August 1972 event which could lead to a high Mach number quasiperpendicular shock and possibly a solar particle event comparable to the Miyake et al. (2012, https://doi.org/10.1038/nature11123) tree ring event.

## 1. Introduction

Of the at least four August 1972 interplanetary coronal mass ejections (ICMEs)/shocks, the event on 4 August has been recognized as having the fastest Sun-to-Earth transit in recorded history, ~14.6 hours (Vaisberg & Zastenker, 1976; Zastenker et al., 1978; D'Uston et al., 1977). The 4 August event had an average speed of ~2900 km s$^{-1}$. The peak of the related solar energetic particle (SEP) fluence was the highest during the space age (Kohl et al., 1973; Van Hollebeke et al., 1974; Levy et al., 1976), an order of magnitude larger than the 1989 and 2003 events (Jiggens et al., 2014). The four magnetic storms that occurred between 2 and 11 August 1972 were only of modest intensities. All four storms could be related to flares/ICMEs associated with their shocks/sheaths. The second of three Forbush decreases that occurred was the largest on record (Pommerantz & Duggal, 1973). The August 1972 overall event was in strong contrast to the Carrington 1–2 September 1859 solar flare/magnetic storm event. Although the Carrington ICME/shock Sun-to-Earth transit time (~17.4 hours; Carrington, 1859) was close to the 4 August 1972 event, the Carrington solar event caused an exceptionally intense magnetic storm of SYM-H/Dst of ~ -1760 nT (Tsurutani et al., 2003, 2023; Lakhina et al., 2012; Lakhina & Tsurutani, 2018), while the 1972 ICMEs did not. None of the 1972 magnetic storms were larger than SYM-H = -154 nT. The Carrington event was not associated with a noticeable SEP event (Wolff et al., 2012). In the above, SYM-H is basically the 1 minute version of Dst (Iyemori, 1990). Dst is a composite measurement of 4 near-equatorial ground magnetometer H components and is a 1 hour average index. The Dst index can quantitatively indicate the total energy of ring current particles (Dessler & Parker, 1969; Sckopke, 1966).

The March 1989 space weather event is in sharp contrast to both the 1972 and 1859 Carrington event, but in different ways. There was an extremely large flare-producing active region in 1989 as was the case with the 1972 event, but the most intense 1989 flares were generally of the low X-class intensities (Allen et al., 1989). On the other hand, a magnetic storm during the 1989 event was the largest SYM-H (-710 nT) event during the space age (Allen et al., 1989; Bolduc, 2002; Lakhina & Tsurutani, 2016, 2018). As far as it is known, the speed of the 1989 ICMEs were no different than ordinary fast (those that produce forward shocks) ICMEs. Tsurutani et al. (2023) have called this a "stealth magnetic storm" because current modeling of a single coronal mass ejection (CME) released from the Sun would not be able to predict the arrival time of the shock/ICME at Earth nor the complexity of the long-duration (~24 hours) massive compound interplanetary event (Lakhina & Tsurutani, 2016).

Although there have been many published reports immediately following the August 1972 and the March 1989 space weather events, it is informative and useful to review the past literature to compare and contrast these two quite different space weather events. Intercomparisons will also be made with the Carrington 1859 space weather event and several intense space age events. It has also recently been shown that extreme solar particle events may have occurred in the distant past (775 CE, 994 CE, 660 BCE, and 7176 BCE) using cosmogenic isotopes ($^{10}$Be, $^{36}$Cl, and $^{14}$C) detected in tree ring and ice core data (Miyake et al., 2012, 2019; Usoskin et al., 2013; Brehm et al., 2022; Koldobskiy et al., 2023). See also Bard et al. (2023) for a possible event at 14300–14299 BP. These events have had peak fluxes that are one or two orders of magnitude higher than either the August 1972 or the March 1989 event (Koldobskiy et al., 2023). See Usoskin et al. (2023) for an excellent review. Can further knowledge about the two space-age SEP events (August 1972 and March 1989) tell us more about what might have caused these more ancient possibly interplanetary particle events? Can the related 1972 and 1989 magnetic storms have been higher than the Carrington event under slightly different conditions? The purpose of this review is to bring together knowledge of ancient extreme solar flares, extreme SEP events and extreme magnetic storms for the readership so that further progress can be made.

In Section 2 we have collected the initial reportage of the August 1972 space weather event and we report the main features of this extreme event with references to the original publications. Section 3 gives more recent results concerning the August 1972 event. Section 4 is a review of the 7–19 March 1989 extreme magnetic storm, the largest during the space age.

Section 5 is devoted to the Halloween 2003 solar flares, presumably the largest flares during the space age. Section 6 discusses the 12 November 1989 particle event, the most thoroughly examined quasi-parallel shock event studied during the space age. Section 7 deals with overshoot particles of supercritical shocks and the formation of seed particles for shock acceleration. The overall summary and conclusions are given in Sections 8 and 9, respectively. Some final comments are given in Section10.

## 2. Initial Reports on the August 1972 Space Weather Event

### 2.1 Solar Flares, CMEs, Forward Shocks and Sudden Impulses (SI$^+$s)$^1$ at Earth

There was significant solar activity during the August 1972 event. There were four major flares at McMath 11976, sunspot group 331. Zirin and Tanaka (1973), using Big Bear and Tel Aviv ground-based imaging observations, reported them to occur on 2 August (at 18:38 UT and 20:05 UT), 4 August (at 06:30 UT), and 7 August (at 15:16 UT). The authors reported a 5$^{th}$ flare on 11 August (at 12:16 UT), but only the late phase of the flare was detected. Most of the other reports on the August 1972 events did not discuss this event, so it will not be covered here. Zirin and Tanaka (1973) reported that "The McMath region showed inverted polarity and high gradients from birth, and flares appear due to strong magnetic shears and gradients across the neutral line produced by sunspot motions." During the 4 August flare, the sunspot group extended to 17° in longitude (Pommerantz & Duggal, 1974). See also Tanaka and Nakagawa (1973), Bhonsle et al. (1976), and Yang et al. (1983) for further solar flare details. Bhonsle et al. (1976) reported Type II and Type IV radio bursts for the four flares. Besides the four X-class flares, there were twenty-seven M-class and fifty-six C-class flares (see also Hajra et al., 2020 for multiple flaring, typical of active regions). Chupp et al. (1973), using OSO-7 observations, reported the first observation of solar gamma ray lines at 0.5 MeV (positron annihilation) and at 2.2 MeV (neutron capture in solar hydrogen). Besides a broad X-ray spectrum they also noted possible gamma ray lines at 4.4 MeV and 6.1 MeV. Zirin and Tanaka (1973) rated the 4 August event a comprehensive flare index of 17, the highest ever observed at that time.

There were presumably CMEs associated with each of the four major flares. Although there were only intermittent interplanetary data just upstream of the Earth, positive sudden impulses (SI$^+$s) detected by magnetometers on the ground clearly indicated the arrivals of fast forward magnetosonic shocks created by the ICME "pistons". Matsushita (1976) indicated that the shocks impacted the Earth at 01:19 UT on 4 August, at 20:54 UT on 4 August, at 14:00 UT on 5 August, and at 23:54 UT on 8 August. These will be called M-timings (for Matsushita, 1976) later in the paper.

Mihalov et al. (1974) examined Pioneer 9 data at 0.78 au for the first four ICME shocks. Pioneer 9 was close to central meridian during the first two flares on 2 August. The date of detection, Alfvénic Mach number and shock normal angle relative to the upstream magnetic field for the four ICME shocks were: 1) 3 August: 4.4, 76°, 2) 3 August: 5.1, 79°, 3) 4 August: 2.9, 64°, and 4) 9 August: 3.8, 37°, respectively. Dryer et al. (1976) estimated the sum of the flare kinetic, magnetic and thermal energies using the solar wind detected at Pioneer 9 assuming a hemispherical 2π solid angle. Their estimates range from $8.8 \times 10^{31}$ to $4.6 \times 10^{32}$ ergs for the four events (further details are given in Table 1). Note that these estimates did not include photon and energetic particle energies. Dryer et al. (1976) also made energy estimates and obtained a slightly higher range of values, $5.8 \times 10^{32}$ to $5.3 \times 10^{33}$ ergs. Smith (1976) and Smith et al. (1977) studied the shocks using Pioneer 9, 10, Explorer 41 and Helios plasma and field data. At Pioneer 10 (2.2 au), Smith detected only one forward shock and one reverse shock. Smith determined that there was no further strong deceleration of the ICME events between 0.8 to 2.2 au. The first three forward shocks detected at Pioneer 9 had presumably coalesced into one shock at Pioneer 10.

---

$^1$ Some of the old terminology used during 1972 has been updated in this report. The phenomenon Sudden Storm Commencements (SSCs) have been replaced by sudden impulses (SI$^+$s: Joselyn & Tsurutani, 1990) since shocks do not necessarily have to be followed by magnetic storms. Driver gases are now called coronal mass ejections (CMEs). Solar energetic particles will be called SEP events, a term not used previously.

The shock at 23:54 UT on 3 August and following sheath on 4 August with the related ICME at Earth received considerable attention. Vaisberg and Zastenker (1976), using the highly elliptical Prognoz Earth-orbiting satellite, determined a solar wind speed of > 1700 km s$^{-1}$ at 1 au for this 4 August event. Their instrument was unfortunately saturated, so the maximum value was not measured.

### 2.2 The Earth's Magnetosphere, Auroras, Substorms and Magnetic Storms

Cahill and Skillman (1977) used Explorer 45 data at 5.2 $R_E$ to show that the magnetopause moved inside the spacecraft orbit several times "when the solar wind velocity and interplanetary magnetic field were high." "There were also times when the compressed field decreased by flux erosion. ~20 nT pulsations in the field magnitude with a ~2 minute period were observed during the first approach to the magnetopause. They appeared to not be surface waves propagating in the magnetopause." These measurements of course do not tell us what the minimum magnetopause subsolar distance was during this event. It is interesting to note that had GPS satellites been operational at the time, they might possibly have been exposed to the solar wind when they crossed the dayside equator at $L \sim 4.2$.

Kawasaki et al. (1973) reported, using high time resolution ground magnetometer data, that on 4 August there were two SI$^+$s, one at 01:19 UT (~+30 nT) and a second at 02:20 UT (~+40 nT). "A storm main phase decrease was brief. It began at ~03:30 UT and by 07:00 UT the recovery phase had already begun. The second storm began at 20:54 UT with a SI$^+$ before the first storm had fully recovered." "It is possible that this storm was associated with the flare on 4 August, implying a transit time of ~14 hours." "This may not be too unrealistic considering the large solar wind velocities observed. Kawasaki et al. (1973) suggested that the transit time for a geomagnetic storm occurring immediately after another is usually much shorter than for an isolated storm." Akasofu (1974) noted that "there was a sudden brightening of the equatorward horizon at 2054 UT on August 4, which coincided exactly with the time of the sudden impulse of the August 4–5 magnetic storm. It appeared slightly above the equatorward horizon at 2104 UT and moved rapidly poleward, reaching the zenith at 2113 UT. It is likely that this particular activity was associated with the substorm that was triggered at the time of the SI$^+$." There was a red aurora associated with the magnetic storm. The red aurora was present at 15:30 UT (local magnetic noon), and it was associated with a substorm that was suddenly intensified at about 14:15 UT which was the most intense substorm (~3000 nT) throughout the entire August event".

Loomer and van Beek (1972) reported that one Canadian auroral zone station detected magnetic deviations as large as ~6000 nT.

Kawasaki et al. (1973) indicated that the third magnetic storm began with a double SI$^+$ event, one at 23:54 UT on 8 August, and the other at 00:36 UT on 9 August. They speculate that this storm was probably related to the flare on 7 August.

**2.3 Solar Energetic Particles (SEPs)**

There are two types of SEP events, 1) those that are formed at the Sun by solar flare related processes (Freier & Webber, 1963; Kallenrode, 2004), and 2) those that are accelerated at forward and reverse shocks in interplanetary space (McDonald et al., 1976; Pesses et al., 1979; Tsurutani et al., 1982) associated with corotating interaction regions (CIRs; Smith & Wolfe, 1976) or with forward shocks associated with fast (> 500 km s$^{-1}$) ICMEs (Tsurutani & Lin, 1985; Kahler, 1992; Cliver, 1996; Cliver et al., 2020). See an extensive review of SEP acceleration and propagation by Kallenrode (2004). Reames (1998) has argued that the most intense SEPs detected at 1 au are due to fast forward ICME shocks. If one has interplanetary energetic particle data, it is relatively easy to separate SEPs accelerated at the flare site (prompt events) from those accelerated at ICME shocks (gradual or delayed events). The SEPs accelerated at the flare site will arrive at Earth within an hour of the flare onset (without an accompanying shock), whereas ICMEs and their shocks can take 1 to 4 days to reach 1 au (Tsurutani et al., 2009). For the latter case, particle flux maxima are typically located at or very close to the ICME forward shock. This will be discussed further in the data analysis section.

For interplanetary shock particle acceleration, there are two fundamental mechanisms typically cited in the literature: particle gradient drift along the shock surface in the direction of a motional electric field (Hudson & Kahn, 1965; Sarris & Van Allen, 1974; Pesses et al., 1979; Decker, 1981) and Fermi acceleration by wave-particle scattering in the upstream and downstream regions relative to the shocks (Axford et al., 1977; Blandford & Ostriker, 1978). The gradient drift mechanism is most effective for quasi-perpendicular shocks and the Fermi acceleration mechanism most effective for quasi-parallel shocks. By quasi-perpendicular shocks is meant that the shock normal subtends a large angle relative to the upstream magnetic field direction, and by quasi-parallel, is meant that the shock normal subtends a small angle relative to the upstream magnetic field direction. An excellent example of quasi-parallel shock particle acceleration is given in Kennel et al. (1984a, 1984b). Examples of quasi-perpendicular shock acceleration are given in Tsurutani and Lin (1985).

Van Hollebeke et al. (1974) made the IMP 4 and 5 SEP 1972 event particle data available in a Goddard Space Flight Center catalogue. Kohl et al. (1973), using IMP 5 and 6 > 10 MeV, > 30 MeV and > 60 MeV proton data, reported energetic particle increases on 2 and 3 August, and speculated that these were due to the 2 August flares. "The peak flux during the extended event interval occurred in early 4 August about the same time as a SI$^+$ occurred" (there was a second major event that began at 15:33 UT on 7 August). Pomerantz and Duggal (1973, 1974) reported a ground-level enhancement (GLE) event (caused by relativistic particles). This initially left a large mystery of the mechanism for the particle acceleration and transport. Levy et al. (1976) modeled this relativistic particle event as first-order Fermi acceleration between two converging fast-forward shocks (this is a third shock acceleration mechanism not mentioned above). The faster shock was initially believed to possibly be a blast wave with a speed of > 5500 km s$^{-1}$ near the Sun (Dryer et al., 1972), but D'Uston et al. (1977), using Prognoz solar wind data, later identified CME material several hours behind the shock, indicating that the shock was driven. The average Sun-Earth speed of the shock was ~2900 km s$^{-1}$ (Pommerantz & Duggal, 1974). The slower shock in the Levy et al. model had an average speed of ~1400 km s$^{-1}$ with a speed of 500 to 700 km s$^{-1}$ at 1 au (Dryer, 1974). This 4 August particle event is the most intense in the history of the space age.

Pommerantz and Duggal (1973) reported three Forbush decreases (Forbush, 1938) during the 1972 August active interval. The second Forbush decrease on 4 August "reached 27% in about 24 hrs". This is the largest Forbush decrease on record. The second most intense event was in July 1959 where the peak decrease was 24%.

### 2.4 Space Weather Effects and Ionospheric Phenomena

Anderson et al. (1974) reported an outage of the Plano Illinois to Cascade Iowa link of the ATT L4 cable at 22:40 UT on 4 August. They calculated that an electric field > 6.5 V km$^{-1}$ would be necessary to cause the outage. At that time the magnetopause was pushed in to distances inside $L$ = 6.6. Using a Plano Earth resistivity model they estimated a geomagnetic disturbance of ~700 nT minute$^{-1}$ which induced a perpendicular surface electric field of ~7.4 V km$^{-1}$. Anderson et al. (1974) concluded: "The geomagnetic disturbances that produced the shutdown were not of the auroral-electrojet type normally associated with disruptions of power systems."

Albertson and Thorson Jr. (1973) reported that "widespread effects on power systems began at approximately 22:42 hours universal time (UT), August 4, 1972 (22:42 UT = 17:42 central daylight time)." "The system operating effects of the storm were more noticeable in northern latitudes and in igneous rock areas" (northern Minnesota, North Dakota, Manitoba, Newfoundland and portions of Ontario and Quebec). Newfoundland and eastern North Dakota seemed to have deviations that were far more severe than normal operating conditions. "The voltage collapsed down to 64% on the Grand Forks end of the Manitoba-North Dakota interconnection." Lincoln and Leighton (1973) reported that there were intense polar cap absorption (PCA) events for long durations blocking all radio communication to the Arctic and Antarctic regions. On 4 August quasi-DC ground currents of ~200 A entered power system transformers and transmission lines causing difficulties in the normal power supply in the USA and Canada.

Mendillo et al. (1973) reported that the solar flare on 6 August caused an ionospheric total electron content increase near 10:15 eastern standard time (EST) of approximately $4.3 \times 10^{16}$ electrons m$^{-2}$ (~30% of the pre-flare value).

# 3 More Recent August 1972 Event Results

## 3.1 Energetic Proton Events and Interplanetary Shocks

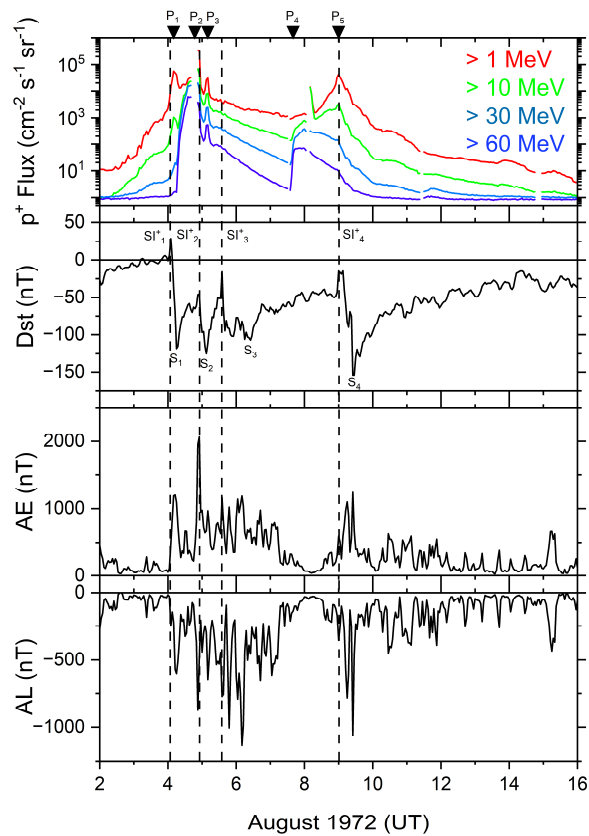

**Figure 1**. The IMP-5 > 1 MeV, > 10 Mev, > 30 MeV and > 60 MeV energetic proton fluxes and simultaneous ground Dst, AE and AL indices from 2 to 15 August 1972. Five proton peak fluxes (P$_1$ through P$_5$) are indicated at the top of the figure. Positive SI$^+$s are indicated above the Dst panel. The peaks of four magnetic storms (S$_1$ through S$_4$) are indicated in the Dst panel.

Figure 1 shows the interplanetary energetic proton fluxes observed by IMP-5 orbiting the Earth, and the ground geomagnetic indices Dst, AE and AL. It should be noted that for the August 1972 event there are no high time resolution 1-minute SYM-H or wide longitude and latitude SML data coverage available. However, this lower resolution data here is useful to show the general relationships of the geomagnetic activity to the energetic solar protons for this event.

There are five clear peaks in the four proton flux energy channels (> 1 MeV, > 10 MeV, > 30 MeV, and > 60 MeV). These are labeled as $P_1$ through $P_5$ in Figure 1. The fluxes of the > 10 MeV channel are: $P_1$ ~1,059 cm$^{-2}$ s$^{-1}$ sr$^{-1}$, $P_2 \geq$ 67,750 cm$^{-2}$ s$^{-1}$ sr$^{-1}$, $P_3$ ~8,238 cm$^{-2}$ s$^{-1}$ sr$^{-1}$, $P_4$ ~794 cm$^{-2}$ s$^{-1}$ sr$^{-1}$, and $P_5$ ~3,500 cm$^{-2}$ s$^{-1}$ sr$^{-1}$. The $P_2$ peak flux at ~21:00 UT on 4 August is the highest flux recorded in the space age. The SEP event would have posed severe health threat to astronauts at the time if they had been en route to the moon, or outside walking on the moon (Lockwood & Hapgood, 2007). Fortunately the event occurred between Apollo missions 16 and 17.

The Dst panel shows four very clear SI$^+$ events. These are denoted by vertical dashed lines and occur at ~02:00 UT (01:19 UT M-timing) on August 4, ~22:00 UT (20:54 UT M timing) on 4 August, ~14:00 UT (14:00 UT M-timing) on 5 August, and ~00:00 UT (23:54 UT 8 August M-timing) on 9 August. The hourly averaged SI$^+$ intensities are: +30 nT, +20 nT, +35 nT, and +28 nT, respectively. SI$^+$s are caused by the high-density plasma sunward of the forward magnetosonic shocks suddenly compressing the magnetosphere (Araki, 1993; Tsurutani & Lakhina, 2014). The large amplitude SI$^+$s detected in the Dst indices indicate that all four shocks were strong, and caused substantial plasma compressions in their downstream regions. Two of the proton flux peaks correspond to SI$^+$s: $P_1$ at ~04:00 UT on 4 August and $P_5$ at ~00:00 UT on 9 August.

Following the four SI$^+$ events are four magnetic storms. The peaks in the storms (minima in Dst) are indicated by $S_1$ through $S_4$, and have the following peak hourly intensities of -118 nT, -125 nT, -107 nT, and -154 nT, respectively. These magnetic storm intensities are "intense", Dst $\leq$ -100 nT (Tsurutani et al., 1988), but not of superstorm intensities, Dst $\leq$ -250 nT (Tsurutani et al., 1992b; Echer et al., 2008a; Meng et al., 2019) or higher. The 1859 Carrington magnetic storm had a SYM-H/Dst intensity of -1760 nT, and the 1989 Hydro Quebec storm a SYM-H intensity of -710 nT.

What is important to note here is the near-simultaneous occurrence of $P_1$ and the $S_1$ shock (represented by SI$^+_1$). $P_1$ occurs ~2 hours after SI$^+_1$. This is consistent with local in situ shock acceleration of the energetic particles and not acceleration at the Sun as in the case of prompt particle acceleration. Also because of the near-simultaneity of the shock and the proton peak flux, the standard shock acceleration mechanisms seem to apply. It is noted that the $P_1$ spectrum is very soft with very little > 30 MeV and > 60 MeV protons present. However these energy protons are present at $P_2$. The latter may be due to either the 2-shock Fermi mechanism proposed by Levy et al. (1976) and Zhao and Li (2014) or shock acceleration of new "seed" particles created by shock $S_2$. Of course both processes may be occurring at the same time.

The $P_2$ energetic proton flux peak occurs slightly upstream of the $S_2$ shock (represented by SI$^+_2$). Levy et al. (1976) modeled this relativistic particle event as first-order Fermi acceleration between two converging shocks, $S_1$ and $S_2$. The faster driven shock $S_2$ had an average Sun-Earth speed of ~2900 km s$^{-1}$, while the slower shock ($S_1$) had an average speed of ~1400 km s$^{-1}$ with a speed of 500 to 700 km s$^{-1}$ at 1 au. They found that the spectrum steepened with increasing energy, γ increasing from a value of 9 or 10 below ~ 0.9 GeV to greater than 20 by 1.2 GeV. Earlier, Bazilevskaya et al. (1973) obtained a spectral index of $\gamma = \sim 10$ for limited GLE data between ~0.5 to 0.9 GeV for this event. See Lockwood et al. (1974) for further comments. Why are these distinctions important ones to make? It is because this 4 August ($P_2$) extreme flux event was different than the standard shock acceleration mechanisms typically attributed to SEPs associated with ICME driven fast shocks mentioned previously (shock gradient drift and Fermi acceleration associated with upstream and downstream plasma waves). Can such a converging shock mechanism be a possible explanation for the ancient extreme energetic particle events? This will be discussed later in the Conclusion section.

$P_3$ occurs ~6 hours after $SI^+_2$. It may be associated with acceleration at shock $S_2$ and the following interplanetary sheath. $P_4$ is not associated with any observable interplanetary feature. $P_5$ is near-coincident with shock $SI^+_4$. It has been previously noted by Tsurutani et al. (1982) that particle peak fluxes and shocks are not necessarily time-coincident. There are many possible explanations. Please see the geometry of the shocks and explanation in the above reference.

### 3.2 Magnetic Storms and Substorms

As mentioned in the above section the August 1972 storm peaks (minima in Dst) of the four magnetic storms were only of moderate intensities: $S_1$ = -118 nT, $S_2$ = -125 nT, $S_3$ = -107 nT and $S_4$ = -154 nT. The storm main phases begin directly after the four $SI^+$s, indicating very short or nonexistent storm initial phases. This indicates that one interpretation is that all four of these magnetic storms are caused by sheath magnetic fields (see Tsurutani et al., 1988; Meng et al., 2019)

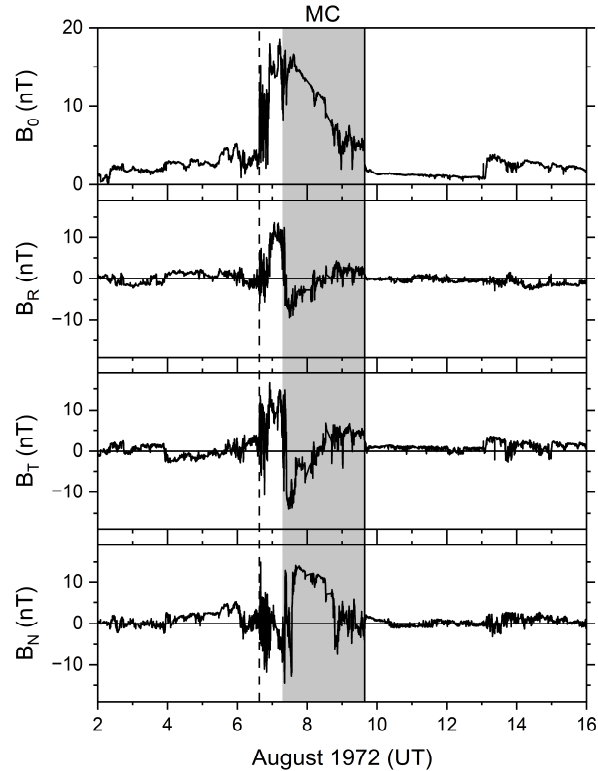

**Figure 2**. The Pioneer 10 interplanetary magnetic field (IMF) magnitude $B_0$ and components $B_R$, $B_T$, and $B_N$ in RTN coordinates. Fast forward and fast reverse shocks are marked by vertical dashed and solid lines, respectively. A magnetic cloud (MC) is denoted by shading.

At the beginning of the August 1972 events, Pioneer 9 at 0.8 au and Pioneer 10 at 2.2 au were in radial alignment. Pioneer 10 was at a solar longitude of -43° relative to Earth. Figure 2 shows the Pioneer 10 magnetic field for these events. The coordinate system used is the spacecraft-centered radial tangential normal (RTN) system where ***R*** points radially away from the Sun towards the point of observation, ***T*** is parallel to the solar equator and is positive in the direction of the orbital motion of the planets, and ***N*** forms the right-hand system (pointing northward).

As noted by Smith (1976), there is a forward shock at ~15:00 UT on 6 August, and a reverse shock at ~15:00 UT on 9 August at Pioneer 10. The forward and reverse shocks are indicated by a vertical dashed and a solid line, respectively in Figure 2. Smith et al. (1977) indicated that the interplanetary magnetic field was greatly evolved while transiting the distance from Pioneer 9 at 0.8 au to Pioneer 10 at 2.2 au. The three forward shocks detected at Pioneer 9 had presumably coalesced into only one forward shock at Pioneer 10. The forward shock speeds at Pioneer 9 and Pioneer 10 were of the order of 600 to 700 km s$^{-1}$, indicating that they had greatly decelerated compared to the near-Sun values.

The Pioneer 10 magnetic field data, although evolved, does give some important information. There is an abrupt discontinuity at ~08:50 UT on 7 August which changes the southward magnetic field orientation to a northward direction. A second discontinuity occurs at ~13:40 UT the same day. After this time the field is large and without waves indicating that this was the MC portion of an ICME (Tsurutani et al., 1988). The large northward IMF on 7 August lacks waves and discontinuities and has a magnitude of ~14 nT in the MC. Smith et al. (1977), using higher time resolution data, indicated a peak magnetic field of ~18 nT. R. P. Lepping (Tsurutani et al., 1992a) analyzed this magnetic event and concluded that it had the form of a highly tilted flux rope with the magnetic field with a northward direction.

Although there were four very intense solar flares, four associated SEP events, and four intense SI$^+$ events (the latter indicating the presence of ICME driven forward shocks) at Earth during the August 1972 event, Pioneer 10 detected only one ICME/MC. How can this be explained? It has been noted in previous studies, such as the international CAWSES space weather effort, that multiple solar flaring from an AR leads to several sheath intervals but only one ICME/MC at Earth/1 au (Tsurutani et al., 2008a). A possible interpretation is that MC-MC interactions (and possible expansions) cause deflections of the MCs in interplanetary space, and thus only one impacts the Earth's magnetosphere. Tsurutani et al. (2008a) has discounted the possibility that some of the shock/sheaths reaching Earth were due to blast waves. It was argued that the other ICMEs were deflected and missed the Earth.

It has also been noted by Echer et al. (2009) that simple ICME propagation models that do not include the possibility of MC-MC collisions and deflections will not give accurate MC arrival times at Earth. Thus, when the Sun has a very active AR causing many intense solar flares and ejecting many fast CMEs, the prediction of magnetic storms associated with solar AR intervals becomes far more difficult.

There was OMNI data for the 4 August 1972 ICME but the magnetic field cutoff was ~44.3 nT with a data gap. The peak magnetic field of ~18.6 nT at 2.2 au would imply a magnetic field intensity of ~87 nT at 1 au assuming a $1/r^2$ magnetic field fall-off with radial distance $r$. This value would be higher if the MC expanded superradially. The estimate of ~87 nT for the 4 August 1972 ICME agrees quite well with the Carrington magnetic field at 1 au estimated minimum intensity of ~90 nT (Tsurutani et al., 2003, 2023). The solar wind velocity transit times of the two ICMEs are comparable.

Cid et al. (2023) have recently reexamined the Pioneer 9 and 10 spacecraft data suggesting that there was an "ICME in a sheath" and the presence of a high-speed stream (HSS) during the region as well.

### 3.3 The Cause of Geomagnetically Induced Currents (GICs)

There has been a new interpretation of the Anderson et al. (1974) reporting of the AT&T line outage during the 4 August event. Boteler and van Beek (1999) have performed model calculations that have shown that "the observed model disturbance was too localized to have been caused by magnetopause currents. Contour plots of the disturbance are instead consistent with an ionospheric current as the source. Equivalent current plots of the electrojet was responsible for the magnetic disturbance and cable system outage." Boteler (2001) concluded "The blackout of the Hydro-Quebec system on March 13, 1989 was due to an enhancement of a westward substorm electrojet resulting from loading and unloading of energy in the magnetosphere." It is interesting that Tsurutani and Hajra (2021) have recently shown that shock triggered supersubstorms have been primarily responsible for intense geomagnetically induced currents (GICs) at the near auroral zone station of Mäntsälä, Finland. In hindsight the Akasofu (1974) reporting of the ~3000 nT substorm onset with shock triggering on the 20:54 UT on 4 August $S_2$ event is consistent with this picture.

There is evidence that the rapid change in magnetic field dB/dt on the 4 August 1972 is responsible for the detonation of US Navy mines in coastal waters off South Vietnam (Knipp et al., 2018).

## 4. The 6–19 March 1989 Event

The March 1989 event interval contained the largest peak intensity magnetic storm that has occurred in the space age and caused the outage of the Hydro-Quebec power system for ~9 hours. Here we present a review of the 1989 solar, energetic particle, magnetic storm and storm and particle effects. Some brief comparisons will be made to the August 1972 and September 1859 Carrington space weather events. In our Summary and Conclusion section, further comparisons between the three space weather events (and others) will be made.

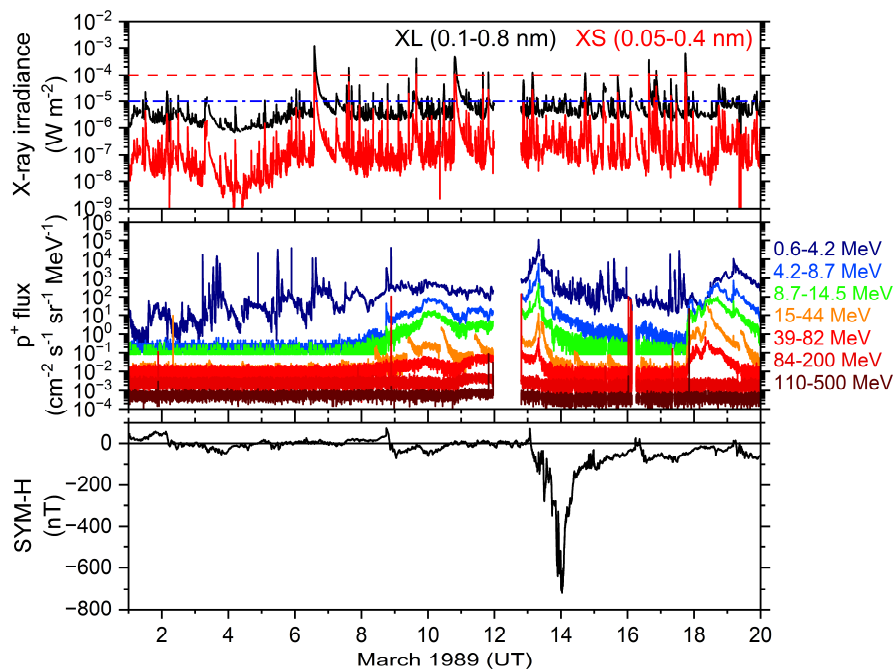

**Figure 3**. From top to bottom, panels show X-ray irradiances at long wavelengths (XL, 0.1–0.8 nm, black) and short wavelengths (XS, 0.05–0.4 nm, red), 0.6–500 MeV proton fluxes, and SYM-H index during 1–19 March 1989. Red dashed and blue dash-dot horizontal lines in the top panel indicate thresholds for X-class and M-class flares, respectively.

Allen et al. (1989) reported most of the solar, interplanetary and magnetospheric effects of the March 1989 Space Weather event. There were radio blackouts, Forbush decreases and other features similar to the August 1972 event. Below we will give the main points taken from their work relative to the topics of this review. For references to the sources of the observations, we refer the reader to the original Allen et al. paper.

### 4.1 Solar Flares

Active region 5395 first appeared on 6 March 1989 at the east limb of the Sun. That was the source of flares for two weeks when the active region was visible from Earth. A particularly large X15/3B flare occurred at 35° N, 69° E at ~14:06 UT on 6 March. Between 6 and 19 March, the AR produced 11 X-class and 48 M-class flares. "The long-lived AR produced one or more flares daily near or surpassing the X1.0 level. The X4.0 flare at 1532 UT on 9 March was a 4-Bright (4B), the highest category in both area and intensity. At ~19:21 UT on 10 March, a flare of intensity X4.5/3B occurred at 22° E longitude."

Comments: The largest X15/3B 6 March flare of the March 1989 interval did not cause a geomagnetic storm at Earth, shown in Figure 3. The flare was an east limb event and was not expected to be geoeffective. However it was noted recently in Meng et al. (2019) that it is possible for east limb flares/ICME shocks/sheaths to indeed be geoeffective. There were more X-class flares during the March 1989 event than during the August 1972 event.

Boteler (2019) has argued "the evidence shows that the disturbance (the large SYM-H = -710 nT magnetic storm) was caused by two CMEs, the first associated with an X4.5 flare on 10 March and the second linked to a M7.3 flare on 12 March. The arrival of the interplanetary CME shock fronts caused SI$^+$s at 01:27 and 07:43 UT on 13 March (see also Allen et al., 1989, below). The transit time and speed of the first (second) interplanetary CME shocks are 54.5 hr (31.5 hr) and 760 km/s (1,320 km/s)".

### 4.2 Solar Energetic Particles

Allen et al. (1989) have stated: "A solar proton event was in progress on 13 March, the launch date of Discovery. However the particle spectrum was "soft", that is, there were relatively few higher-energy (> 30 MeV) protons and alpha particles present. The softness of the particle spectrum prevented a radiation hazard to the mission. The Soviet MIR which had a maximum orbital latitude of ~51° was expected to be exposed to the interplanetary radiation but luckily was not exposed to a major hard spectral event".

### 4.3 Geomagnetic Storms and Substorms

Allen et al. (1989): "This storm had one of the most disturbed 24 hr periods of any since the mid-19$^{th}$ century". "The Ap* value ranks as the third largest magnetic storm since 1932". "According to the AA* index this was the largest magnetic storm since 1868 as recorded by the 2-station network of antipodal sites in the U.K. and Australia." "The storm started with a SI$^+$ at 0128 UT on 13 March. Another SI$^+$ occurred at 0747 UT the same day. Kamei et al. (1989) showed that the AE index ranged from values near zero to as large as ~3,000 nT. It should be mentioned that many of the AE stations were saturated".

Comment: The SYM-H = -710 nT magnetic storm at 00:46 UT on 13–14 March was the largest peak intensity magnetic storm that has occurred during the space age. The magnetic storm is particularly unusual because of its long duration. The storm onset occurred at ~01:31 UT 13 March with a SI$^+$ of magnitude +72 nT. Thus, the duration of the storm main phase was 23 hours and 15 minute long. The typical magnetic storms of the 10 events selected by Gonzalez and Tsurutani (1987) of Dst ≤ -100 nT was ~12 hours (see the much larger storm selection by Echer et al. (2008b) giving approximately the same result).. What caused this 1989 storm main phase to be so unusually long in duration?

The interplanetary cause of the 1989 magnetic storm was discussed by Boteler (2019, above) and earlier by Lakhina and Tsurutani (2016) based on an examination of the profile of the Dst indices for the storm. The interplanetary event was a compound one. There were several interplanetary sheaths and at least one magnetic cloud that comprised the massive interplanetary structure that created the extremely long magnetic storm main phase, presumably through interplanetary southward magnetic field components. This is in agreement with the finding by Allen et al. (1989) of two SI$^+$s (caused by forward shocks, followed by sheaths). Thus the interplanetary event that caused this very long and intense magnetic storm was associated with multiple solar flares, two shock/sheaths and at least one MC. The solar wind energy input into the ring current was intense and of long duration.

This long duration storm indicates that not only the peak intensity should be considered in assessing its magnitude/importance, but also its duration. As particles are injected into the magnetospheric ring current, they are continuously being lost to the ionosphere and atmosphere at the same time (Kozyra & Liemohn, 2003; Liemohn & Kozyra, 2003). The primary loss processes are convection out the dayside magnetopause and the secondary, charge exchange. Losses due to Coulomb collisions and wave-particle interactions are neglible in comparison. For highly energetic ring current particles in extremely large convection events, they may be lost in ~2 to 10 hours (Kozyra & Liemohn, 2003; Liemohn & Kozyra, 2003). Note that the above two studies were performed before even faster convection events such as the Carrington storm was known.

The extremely long duration of the 1989 magnetic storm and its very high peak intensity implies that more ring current energy is being injected into the magnetosphere than a typical ~12 hour main phase storm Dst < -100 nT (the 10 storms studied by Gonzalez and Tsurutani, 1987 average main phase duration was ~12 hours). To model the total ring current particle energy for magnetic storms, the proper scales for the two main processes must be taken into account. It should be noted that the loss rates are minimum *L* and convection electric field dependent (Liemohn & Kozyra, 2003; Tsurutani et al., 2018), so the modeling should be done carefully. Is it possible that the 1989 storm ring current total energy was larger than the Carrington short-duration main phase storm ring current total energy? Possibly so.

A similar misperception takes place when intercomparing geomagnetic activity for the different solar cycle phases. Magnetic storms typically occur during periods when there is a maximum of sunspots, or solar maximum. However during the declining phase of the solar cycle, dominated by solar coronal holes and high speed solar winds, less intense auroral zone geomagnetic activity known as high-intensity long-duration continuous AE activity (HILDCAAs: Tsurutani & Gonzalez, 1987; Hajra et al., 2013) occurs. Clearly magnetic storms have higher energy per time inputs than do HILDCAAs. However HILDCAAs inject energy continuously into the magnetosphere and it has been argued that it is possible that more energy is input into the magnetosphere during the declining phase than during solar maximum (Tsurutani et al., 1995, 2006; Kozyra et al., 2006; Turner et al., 2006; Hajra et al., 2014a). It has been noted that relativistic magnetospheric electrons are detected predominately during this phase of the solar cycle (Baker et al., 1990; Li et al., 2001; Hajra et al., 2014b). Hajra et al. (2015, 2024) have indicated that the acceleration of these relativistic electrons are associated with chorus (Horne et al., 2005; Thorne et al., 2013) during HILDCAA events.

### 4.4 Magnetopause Location

Allen et al. (1989) showed that the magnetopause moved inside geostationary orbit ($L = 6.6$) during 13 and 14 March using GOES 7 and GOES 6 magnetic field observations. By modeling, the authors determined a minimum subsolar location of 4.7 $R_E$. They calculated that the work done by the solar wind was $4\times10^{22}$ ergs, about 1/6 of the US daily electrical energy consumption in 1987. The extreme inward magnetopause movement during 13 March is similar to that of the August 1972 event.

### 4.5 Effects on Space Systems, Communications and Navigational Difficulties

Allen et al. (1989) report numerous problems with satellites, communication and navigational difficulties throughout the whole 6–19 March interval (and beyond). Here we will cite only a few space anomalies noted by the authors. "GOES 7 had a communications circuit anomaly on March 12 and lost imagery". "The system also had a communications outage (ground control problem) on March 13." "Japanese geostationary communications satellite CS-3B had a severe problem at 1050 UT on March 17 that involved failure and permanent loss of half of the dual redundant command circuitry onboard". "The ESA MARECS-1 satellite had many switching events on March 3, 17 and 29". "The NASA SMM satellite dropped a half km at the start of the March 13 storm and dropped perhaps 3 miles during the entire disturbed period."

### 4.6 Ionospheric Effects

Chakraborty et al. (2008) have detected "abnormal height rise and corresponding drop of plasma density at the equatorial F-layer in the post midnight period of both the main and recovery phases of the 13-14 March 1989 storm". These features maybe related to storm convection electric field features (Tsurutani et al., 2008b). Tsurutani et al. (2012) have modeled the Carrington magnetic storm assuming the convective electric fields derived in Tsurutani et al. (2003). The authors found that the uplift of the ionospheric oxygen ions would be quite substantial on the dayside during the storm main phase. A substantial increase in drag for low altitude satellites would occur if a Carrington storm occurred today.

### 4.7 Power Failures and Intense Auroral Currents

"The Hydro-Quebec power system was out for ~9 hrs during the 13 March magnetic storm. A "large ambient magnetic field" change at 0244 local time was the cause of the outage. After the transformers became saturated by line harmonics, compensators tripped circuit breakers shutting down ~44% of the distributed power. When four other compensators shut down, the system crashed" (Allen et al., 1989).

Allen et al. (1989) also stated: "There was a simultaneous power loss (within one second) on 6 different 130 KV lines in central and southern Sweden at the same time as the Hydro-Quebec outage." "Brilliant auroras were seen across the US during the nights of March12-13 and March 13-14." "Auroras were reported during the London England night of March 13-14, extending to the southern horizon. It was extremely bright in the 630 nm line".

Boteler (2019) indicated that the "second $SI^+$ occurred at the same time as the substorm that impacted the Hydro-Quebec system …. and this caused the production of a larger geomagnetically induced currents that caused the Hydro-Quebec blackout." Zhang et al. (2023) have performed a global MHD simulation of jumps in solar wind ram pressure (forward shocks). "They find that a pair of field-aligned currents result and travel in the anti-sunward direction. Soon afterwards another pair appears and travels in the anti-sunward direction. The horizontal ionospheric current associated with the main impulse remains strong when propagating to the nightside". The authors suggest that the Hydro Quebec blackout could have been caused by the Hall current passing over the HQ power system.

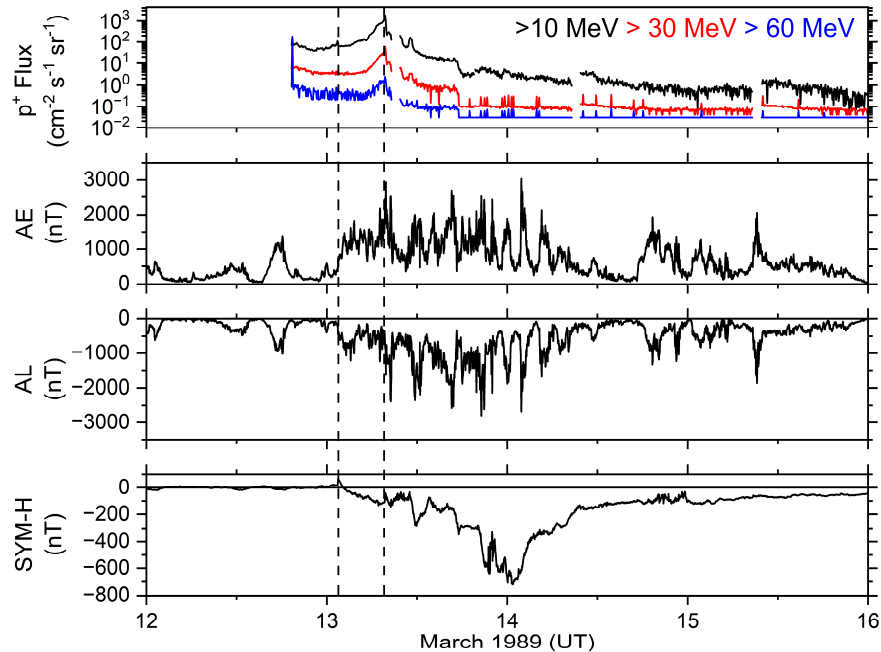

**Figure 4**. From top to bottom the panels are: proton fluxes for > 10 MeV, > 30 MeV and > 60 MeV (5-minute resolution), the AE index, the AL index, and the SYM-H index (1-minute resolution).

Figure 4 shows the peak interplanetary energetic particle event during the March 1989 event interval. The peak flux occurs at ~07:45 UT on 13 March with a > 10 MeV peak flux of 1860 $cm^{-2}$ $s^{-1}$ $sr^{-1}$. This value is an order of magnitude lower than the August 1972 event. The dashed vertical line at ~01:26 UT identifies a $SI^+$ of magnitude +72 nT. There is a second $SI^+$ noted by Allen et al. (1989) which occurred at ~07:40 UT on 13 March. It is indicated by a vertical dashed line. The $SI^+$ had a magnitude of +72 nT (SYM-H change from -112 nT to -40 nT). The $SI^+$ signifies the arrival of a fast forward shock (there were no interplanetary 1 au data available). In Figure 4 the peak of the interplanetary particle event is time coincident with the $SI^+$.

## 5 Halloween 2003 Event

The "Halloween" event of October–November 2003 was associated with large flarings from a solar active region (AR 0486). Tsurutani et al. (2005, 2009) reported the unsaturated EUV flare intensities for 28, 29 October, and 4 November 2003. The 28 October flare was the largest EUV flare in recorded history. The NOAA X-ray detectors were saturated for two of the flares (28 October, and 4 November) and no direct measurements for the peak intensities were possible. However, estimates were made by extrapolation. The 4 November flare was estimated as X28 and the 28 October flare estimated as X17. Thomson et al. (2004) used VLF wave phase change measurements to estimate that the 4 November flare had a magnitude of X45 ± 5. The 29

October X-ray flare was not saturated in the NOAA detectors and was rated as an X10 intensity. The magnetic storms associated with the three flares/ICMEs were Dst = -353 nT on 29–30 October 2003 due to MC fields (Skoug et al., 2004; Mannucci et al., 2005; Alex et al., 2006), -406 nT on 30–31 October 2003 due to sheath + MC fields (Alex et al., 2006; Meng et al., 2019), and no storm on 6–8 November (the flare was a limb flare). It is noted that the ICME speeds for the 28 and 29 October events were reasonably fast but not nearly as fast as the August 1972 and Carrington 1859 events. The 28 October flare caused a ~25 TECU increase (1 TECU or TEC unit = $10^{12}$ electrons cm$^{-2}$; TEC represents ionospheric total electron content) or ~30% above background increase in electron column density (Tsurutani et al., 2005). This is equal to the largest TEC enhancement on record (see Mendillo et al., 1973 for the 6 August 1972 flare).

The 28 October and 4 November 2003 flares were mentioned because they may have been comparable or greater in intensity than the Carrington and August 1972 flares. However the resultant magnetic storm peak intensities at Earth were far less intense than either the Carrington or 1989 storms. The Halloween solar energetic particle peak fluences were far less intense than the August 1972 event.

## 6. The 12 November 1978 Quasiparellel Shock and Energetic Particle Event

We discuss the results of the 12 November 1978 quasiparallel shock here to emphasize that the shock acceleration mechanism of Axford et al. (1977) is not the most efficient mechanism. This will be discussed further in the Conclusion section. The International Sun-Earth Explorer (ISEE) science team under the leadership of C. F. Kennel (Kennel et al., 1984a, 1984b) studied the acceleration of energetic particles at a quasiparallel shock that occurred on 12 November 1978. The authors used a full complement of plasma, magnetic field, plasma wave and energetic particle data. This was the most comprehensive study of shock-particle acceleration ever performed. The shock was high speed (612 km s$^{-1}$), supercritical and quasiparallel ($\theta_{kB}$ = 41°) in nature. A Mach number of 2.8 was estimated. The acceleration was due to a Fermi mechanism associated with particle scattering by self-generated waves both upstream and downstream of the shock in accordance with the Axford et al. (1977) scenario.

"The flux of > 35 keV protons increased by a factor of 15 in the last 45 min and 270 Re before shock encounter with the ISEE-3 spacecraft."

One important point of mentioning this event is that the interplanetary quasiparallel shock was not effective in causing the acceleration of MeV to GeV energetic particles. The particle spectra was quite soft.

## 7. Foreshock Particles and Waves: Supercritical Shock Phenomena (Seed Particles)

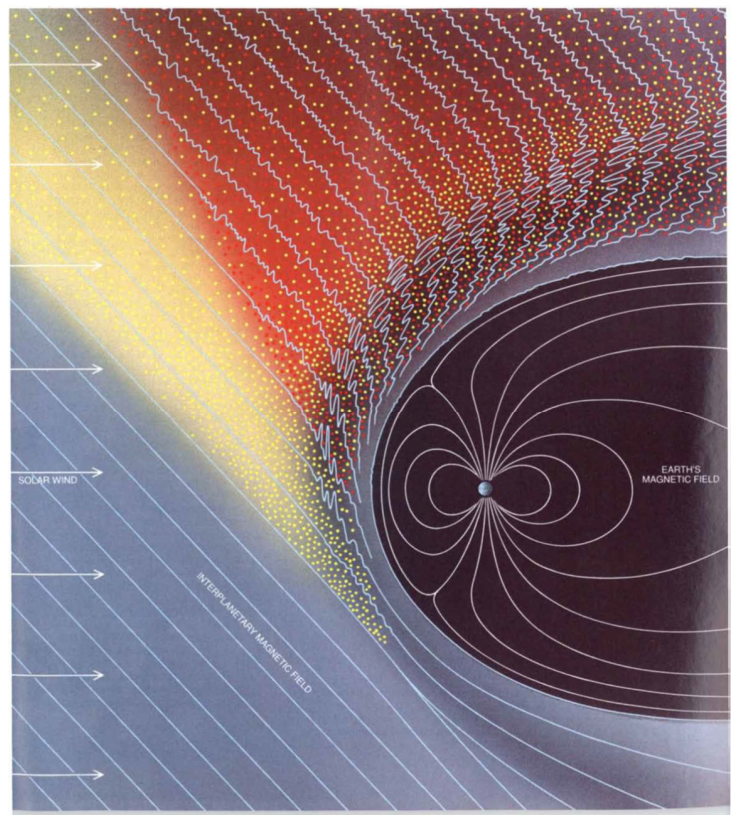

**Figure 5**. The Earth's foreshock. The Sun is on the left out of the picture frame. The solar wind is flowing from left to right. The interplanetary magnetic field lines are at a Parker spiral angle of ~45° relative to the Earth-Sun line. The electron foreshock is shown in yellow and the proton/ion foreshock shown in red. The image is from Sagdeev and Kennel (1991), adapted from Tsurutani and Rodriguez (1981). It should be noted that the Earth's magnetosphere should be rotated by 90° for a correct representation.

The "foreshock" region is the area upstream of collisionless shocks. Figure 5 is a schematic of the Earth's foreshock taken from Sagdeev and Kennel (1991). At the Earth's distance from the Sun, 1 au, the Parker magnetic field lines are approximately 45° relative to the Earth-Sun line. The interplanetary magnetic fields are tangent to the bow shock at ~ 2 pm local time. The shock normal direction is perpendicular to the tangent magnetic field line, so the shock is a perpendicular shock. At ~10 am local time, the interplanetary magnetic field is parallel to the shock normal direction, so the shock is a parallel shock. In between ~2 pm and ~10 am the shock ranges from quasi-perpendicular to quasi-parallel.

Planetary bow shocks in our solar system are high Mach number shocks. The Alfvén wave speed upstream of the Earth's bow shock is ~50 km s$^{-1}$ and the magnetosonic wave speed ~70 km s$^{-1}$ with the solar wind speed varying from ~400 to ~700 km s$^{-1}$. Thus the magnetosonic Mach number for the Earth's bow shock is 7-10 depending on the solar wind speed. The magnetosonic Mach numbers of the Jovian and Saturnian bow shocks are even higher. Neither Alfvén nor magnetosonic waves are able to propagate from the Earth's magnetosphere into the foreshock region because their velocities are well below the solar wind speed. However because the shock is "supercritical" (Sagdeev & Kennel, 1991) shock energy dissipation is partially accomplished by overturned/reflected energetic ions and electrons. These are the particles shown in yellow (electrons) and red (protons) in the foreshock region. The particle energy density is much less than the interplanetary magnetic field energy density. Thus because of the relatively low densities, the particles are guided by the interplanetary magnetic fields. In the figure the majority of particles are found from the quasi-perpendicular portion of the bow shock through the quasi-parallel portion to the parallel portion of the bow shock. Because electrons of the same kinetic energy of protons have much greater velocities, the electron foreshock has a region that is separated from the proton foreshock (Tsurutani & Rodriguez, 1981). See Bonifazi and Moreno (1981a, 1981b), Tanaka et al. (1983), Thomsen et al. (1983a, 1983b) for other mechanisms to create foreshock ions.

The figure also shows wavy lines particularly in the proton foreshock region. The characteristics of quasi parallel and parallel shocks are broad regions of large amplitude waves (Kennel et al., 1984a, 1984b). The waves present far upstream of the bow shock are due to local generation of the waves by an ion-beam instability (Greenstadt et al., 1968; Asbridge et al., 1968; Russell et al., 1971; Lin et al. 1974; West and Buck, 1976; Rodgriguez and Gurnett, 1976; Tsurutani et al., 1993). There are also whistler mode waves generated by beaming electrons (Tsurutani et al., 2001), but these are not shown in the schematic. It should also be mentioned that whistler mode waves can have speeds that are higher than the solar wind speed. So it is possible that this mode of plasma waves can propagate from the magnetosphere into the upstream region. However whistler mode waves have not found to be prominent in the Earth's foreshock region.

The International Sun Earth Explorer (ISEE) NASA-ESA mission extensively studied the Earth's foreshock region with their 3 spacecraft (ISEE-1, ISEE-2 and ISEE-3). Their initial results were published in a JGR special issue (April, 1981). We suggest interested readers to study these articles for previous foreshock wave, particle and instability results.

The foreshock of Jupiter has also been studied (Tsurutani et al., 1993) but with less thoroughness than that of the Earth. Jupiter is ~5 au from the Sun. Because the solar wind plasma decreases as ~1/r$^2$ and the magnetic field less rapidly (Smith, 1979), the Jovian bow shock Mach number is ~10–15. It is notable that while the Earth's foreshock is populated by 30–50 keV ions, the Jovian foreshock has MeV ions (it should be mentioned that most of these very energetic ions could escape from the magnetosphere (Simpson et al., 1974) and not simply supercritical shock dissipation particles.

In both the Earth's and Jupiter's foreshocks all reports of particle streaming have been in the antiplanetary direction. There have been little or no reports of streaming back towards the bow shock. At least not in substantial fluxes.

Comment: There has been a debate about the sources of the "seed particles", those particles that are accelerated to MeV energies by shocks. Are they the lowly solar wind protons, or are they the foreshock particles of supercritical shocks as mentioned by Sagdeev and Kennel (1991) above? Another possibility is that particles accelerated by one ICME shock are further accelerated by a second or third shock (Tsurutani et al., 1982; Gopalswamy et al., 2002). Tsurutani et al. (2020) have mentioned that the seed particle question is one of the biggest problems in doing computer modeling of shock acceleration of particles to MeV energies.

A second point of note is that ICME shocks have a similar convex shape as planetary bow shocks. Thus part of the ICME shocks are quasi-parallel in nature and part quai-perpendicular in nature. Particles are accelerated in the two different regions by different physical mechanisms. Thus there is additional complexity in the problem of shock-particle interactions. This problem is in addition to the difficulties mentioned in Reames (2023).

## 8 Summary

### 8.1 Active Regions and Solar Flares

The 1972, 1989 and 1859 space weather intervals were all associated with large solar ARs. The ARs caused continuous multiple flares with a range in intensities (several X-class, many more M-class and the most C-class). The estimates of the 1972 flare energies made from plasma data measurements are comparable or larger than the flare intensity estimates from X-ray fluxes. Thus one must worry that the X-ray derived flare energies are substantial understimates of the total flare energy if one includes plasmas and magnetic fields and energetic particles. Of course all types of energy estimates are order-of-magnitude and readers should take this into consideration. It is impossible with currently limited ground and space based instrumentation to get more accurate estimates.

It is clear from this review that the features of space weather are not necessarily linearly related to each other. The largest solar flares are not necessarily related to the largest SEP events nor the largest magnetic storms. And vice versa.

### 8.2 Energetic Interplanetary Particles

The 1972 peak particle event was the largest recorded in the space age (Jiggens et al., 2014). It was associated with a Fermi acceleration mechanism with particles mirroring between two fast-forward shocks. The particle energy spectrum was soft. The 1989 particle event also had a soft spectrum where the $E > 10$ MeV peak flux was an order of magnitude lower than the 1972 event. The 1989 particle event was time-coincident with a strong interplanetary shock so the acceleration mechanism would have been either gradient drift/electric fields or the Axford et al.

(1977) wave-particle scattering mechanism. Since the 1859 Carrington event occurred so long ago, it is not possible to have direct measurements of any related particle event. Wolff et al. (2012) did not find substantial particle flux evidence in the ice core samples studied. If there was a related particle event, the spectrum must have been soft. We have mentioned that the Kennel et al. (1984a, 1984b) study of particle acceleration at a quasiparallel shock by wave particle scattering indicated a very soft energy spectrum with a lack of MeV energy protons.

### 8.3 Magnetic Storms

There were four major (Dst < -100 nT) magnetic storms created during the 1972 space weather interval. The maximum event was Dst = -154 nT. Each storm could be related to fast ICMEs/shocks/SI$^+$s and were most probably caused by sheath $B_s$ (IMF southward component) fields. Pioneer 10 data was used to identify the largest magnetic field magnitude detected during the interval as a magnetic cloud. The cloud had a northwardly directed $B_z$. The major storm of the 1989 interval was exceptionally long (23 hours and 15 minutes) and was caused by $B_s$ fields in at least two sheaths and one MC. This exceptionally long magnetic storm indicates that the ring current energy was being pumped into the magnetosphere for a much longer time than either the Carrington storm or the 1972 storms. Using a different measure (the total ring current energy over time), could the 1989 storm have been larger than the Carrington event? Possibly so if the decay constants were ~2 to 10 hours as suggested by Kozyra and Liemohn (2003). This was a compound interplanetary event, in contrast to the simpler events of 1972. The 1989 storm was a "stealth" event because prediction of the time and intensity of the storm at Earth is not yet possible with current codes. In honor of the first paper published on this space weather event we suggest calling this 1989 event the (Joseph) Allen magnetic storm.

The southward component of the Carrington 1859 storm has been estimated to be 90 nT at 1 au. The northward component of the 1972 CME has been estimated as > 87 nT. The two numbers are comparable, only the directions of orientation are opposite. The speeds of the 1972 fastest CME at 1 au was slightly faster than the 1859 CME.

## 9. Conclusions

It has been well recognized that extreme space weather events are associated with large solar active regions which continuously flare during their lifetimes. This is no different from the examples presented here. However from this review it is quite evident that the flaring from an AR is stochastic in nature, sometimes creating a C-class flare (most often) and more rarely creating a X-class flare. Perhaps the largest flare in recorded history is the 4 November 2003 flare with estimated intensity ranging from X28 (extrapolated) to X45 ± 5. This was, however an east limb flare which was not particularly geoeffective. The 28 October 2003 flare was the largest EUV flare in recorded history with an extrapolated X-ray value of X17. The energetic particle fluences from these three events were substantial, but below the values of the 1972 event and the 1989 event. The magnetic storm peak intensities associated with the flare/ICMEs of the 28 and 29 October 2003 events were substantially more intense than the four 1972 storms but lower than the 1989 storm. Table 1 summarises the salient features of the principal space weather events discussed in the paper and provides a quick comparative view of our findings. It is clear that there are no one-to-one links between flare energy, CME speed, energetic particle fluences and magnetic storm intensity.

Furthermore, it is found that magnetic reconnection at ARs can cause flares of extreme intensities as measured in X-rays, in EUV flux (EUV fluxes have been shown to be sometimes more energetic than X-ray fluxes; Tsurutani et al., 2005; Liu et al., 2024), and in plasmas and in energetic particles. However how this energy manifests itself is clearly varied. At times great kinetic energy is expended primarily in photons or in extremely fast ICMEs and at other times in the acceleration of MeV and GeV protons. This review of the literature has told us that this happens but not the reason why.

**Table 1. Salient features of some of the major space weather events**

| S W Event | Solar Flare :class :energy (ergs) | Magnetic Storm Dst/Sym-H (nT) | >10 MeV $H^+$ flux ($cm^{-2}$ $s^{-1}$ $sr^{-1}$) | $SI^+$ event (nT) | CME Transit time (hours) | Remarks |
|---|---|---|---|---|---|---|
| Carrington, 1859 | X46 to X126 [Ha] ~$10^{32}$ [H] | -1760[T] | Not seen | +120[T] | 17.4[C] | Most intense magnetic storm peak intensity in recorded history |
| 2 Aug 1972 (18:38 UT) | ~$8.8 \times 10^{31}$[D] | -118 | 1,059 | +30 | | |
| 2 Aug 1972 (20:05 UT) | ~$2.6 \times 10^{33}$ [V] | | | | | |
| 4 Aug 1972 (6:30 UT) | X20 ~$10^{32}$ [D] ~$5.3 \times 10^{33}$ [V] | -125 | 67,750 | +20 | 14.6[V] | Fastest CME transit in recorded history. Weak magnetic storm. Fermi accn of particles with 2 ICME shocks |
| 7 Aug 1972 (15:16 UT) | ~$3.8 \times 10^{32}$ [D] ~$5.8 \times 10^{32}$ [V] | -107 | ~$3 \times 10^3$ | +35 | | |
| March 1989 | X4.0 (9 Mar) X4.5 (10 Mar) | -710 | 1,860 | | | Largest magnetic storm in space age. Storm main phase (23 hours |
| | | | | | | and 15 minutes) longest in recorded history. Compound interplanetary event. |
| Halloween (2003) | X17 (28 Oct)[(N*,A)] X10 (29 Oct)[(A)] X28 (4 Nov)[(N*)] X45 (4 Nov)[(To)] | -353 -406 | 29,500 | ~+60 ~+45 | 19 19 | Although the 28 Oct flare was extrapolated by NOAA to an intensity of X17, this was the largest EUV flare in recorded history. The EUV energy was more than the estimated X-ray energy[(T5)] |
| Miyake 775 AD | $2 \times 10^{35}$ [(M)] | | $10^2$ to $10^3$ times > Aug 1972. Hard spectra | | | |

**Notes**. H = Harvey (in T2003); Ha = Hayakawa et al. (2023); D = Dryer et al. (1976); V = Vaisberg & Zastenker (1976); T = Tsurutani et al. (2003); C = Carrington (1859); M = Miyake et al. (2012); T5 = Tsurutani et al. (2005); N = NOAA; N* = NOAA extrapolated; To = Tomson et al. (2004); A = Alex et al. (2006).

### 9.1 Particle Acceleration at Fast Shocks

There are different types of "Fermi acceleration" at shocks, some of them discussed in this review. The Axford et al. (1977) picture of upstream and downstream waves acting as the "reflection walls" appears to be the weakest Fermi mechanism as was shown in the Kennel et al. (1984a, 1984b) papers. The particle fluxes were modest and the energies below MeV values. Why is this? It is well known from cyclotron resonant interactions, that there cannot be resonance with particles of 90° pitch angles, except for oblique waves at high relativistic energies (Horne et al., 2003). Thus the diffusion across this barrier/wall is likely to be very slow. This "problem" with the construct of the Axford-type of particle acceleration is not taken into account by theory. Foreshock ion beams generate magnetosonic (compressive) waves through an anomalous Doppler shifted cyclotron resonance (Tsurutani et al., 1993). These magnetic compressive waves could cause mirroring of energetic particles. The theory of Fermi acceleration needs futher development to include both mirroring and pitch angle scattering across 90° pitch.

It was shown by Levy et al. (1976) that the largest interplanetary energetic particle in space age history was due to Fermi acceleration between two fast forward shocks (see also Zhao & Li, 2014 for Fermi acceleration between two shocks). Both shocks were propagating away from the Sun, with the one closer to the Sun propagating at a much faster speed. The resultant particle spectrum was soft. See also Gopalswamy et al. (2002) for discussion of multiple ICME shock acceleration of energetic particles.

In the Levy et al. (1976) event the two "Fermi walls" were the shocks. Perhaps an even more efficient Fermi acceleration configuration would be an event between a fast forward shock and a fast reverse shock? This would involve two ICMEs. The first one would have to be strong enough to create not only a forward shock but also a reverse shock. Then a second fast ICME would create the fast forward shock or the closing wall. Unfortunately such a case has not been detected to our knowledge. There is an additional caveat that the reverse shock may be too weak to actually propagate towards the Sun in inertial space. Although the fast reverse shock may be propagating in the direction of the Sun, it may actually be moving anti-sunward due to the solar wind convection. Tsurutani et al. (1982) noted that CIR reverse shocks were associated with higher energetic proton fluxes than for forward shocks. The causes of these observational features are still not understood.

It would be better for Fermi acceleration efficiency to have a "standing wall". We know that the heliospheric current sheet (Smith & Wolfe, 1976) is one such object. An ICME fast shock producing energetic particles could use the heliospheric current sheet (HCS) as the second wall of the accelerator. However, we note that an HCS is always present in the heliosphere and no one has yet reported strong particle acceleration related to HCSs. Perhaps HCSs are a bit diffuse and not good magnetic mirrors?

Planetary bow shocks are standing high Mach number supercritical reverse shocks. One might think that these would be perfect as energetic particle accelerators. Why aren't MeV and GeV particles observed in the Earth's foreshock? The reason might have to do with scale. The Jovian foreshock has more energetic particles than the Earth's foreshock. ICME shock scales are larger still and astrophysical shocks even larger. This is another argument why the gradient B drift/electric field mechanism (Hudson & Kahn, 1965; Sarris & Van Allen, 1977) needs to be considered more seriously as the major mechanism accelerating SEPs.

The Hudson and Kahn (1965) and Sarris and Van Allen (1977) mechanism is clearly taking place at quasiperpendicular shocks (Tsurutani & Lin, 1985). However a drawback to this mechanism to generate extremely large energetic proton fluxes is that close to the Sun the interplanetary magnetic fields are radially directed and ICME shocks going out radially from the Sun would be parallel or quasiparallel in nature. How could a quasiperpendicular shock be generated close to the Sun? One possibility would involve a two event scenario. First a magnetic cloud is released from the Sun. A fast CME is released a short time later. As the shock penetrates the MC, the shock could become quasiperpendicular. However, again, to our knowledge such an event has never been observed. And there is a large caveat. MCs are low beta regions, so the shock entering the MC would eventually become an evanescent wave.

Cliver et al. (2020) have mentioned a more direct mechanism of perpendicular shock formation and acceleration of particles. The authors note that "The longitudes of flares associated with the top third (24 of 72) of GLEs (ground level events) in terms of their > 430 MeV fluences are primarily distributed over E20-W100 with a skew toward disk center." "The above tendencies favor a CME-driven shock source over a flare-resident acceleration process for high-energy solar protons. GLE spectra show a trend, with broad scatter, from hard spectra for events originating in eruptive flares beyond the west limb to soft spectra for GLEs with sources near central meridian." The idea is that the superradial expansion of the solar wind tangent to the radial direction from the Sun will form a perpendicular shock. Interplanetary magnetic connection to west limb flares will allow > 430 MeV particles to reach Earth. Of course east limb flares will also produce perpendicular shocks but there will not typically be a interplanetary Archimedean orthospiral (opposite rotational sense from an Archimedean spiral) magnetic field so the particles could reach the Earth. This idea is a very interesting one to possibly explain the 775 AD event of Miyake et al. (2012). A collosal west limb flare or series of flares might be the source of the extreme particle event.

**9.2 Does this review information change how large magnetospheric electric fields and magnetic storms can be?**

The answer to this question is "yes". Tsurutani and Lakhina (2014) have already modeled a hypothetical ICME with initial speed of 3000 km s$^{-1}$ at the Sun and obtained a possible interplanetary electric field of ~340 mV m$^{-1}$ and a storm intensity of Dst = -3500 nT (assuming that saturation does not occur at Dst = -2500 nT). However, now with the knowledge that for one of the 1972 ICME events that the average transit speed was 2800 km s$^{-1}$ and that the near-Sun speed might have been ~5500 km s$^{-1}$ (Dryer et al., 1974), it is clear that if an AR releases many fast ICMEs to clear out the heliosphere so that an ICME speed near ~5000 km s$^{-1}$ is possible at 1 au, the interplanetary electric field would increase to a value of ~630 mV m$^{-1}$. Such very strong interplanetary electric fields will lead to extreme uplift of the dayside ionosphere (and atmosphere; Lakhina & Tsurutani, 2017) causing severe drag and loss of low altitude satellites (Tsurutani et al., 2007). This electric field and ionospheric uplift would be substantially larger than the Carrington storm as modeled by Tsurutani et al. (2012).

# 10. Final Comments

Tsurutani and Lakhina (2014) had postulated that shock Mach numbers as large as 45 were possible associated with CME speeds of 2700 km s$^{-1}$ near Earth. Assuming the above scenario that multiple fast ICMEs clean out the region between the Sun and Earth and that 5000 km s$^{-1}$ ICMEs are possible at 1 au, shocks with Mach numbers of ~83 are possible. Because the magnetic fields are less radial at 1 au, quasiperpendicular shock acceleration will be possible. Highly energetic overturn/reflected particles would serve as the seed particles for gradient drift acceleration. Thus it is possible that the August 1972 solar event could have produced a Miyake et al. (2012) 775 AD hard spectral particle event if the interplanetary conditions were a bit different.

**Acknowledgments**
The work of R.H. is funded by the "Hundred Talents Program" of the Chinese Academy of Sciences (CAS), and the Excellent Young Scientists Fund Program (Overseas) of the National Natural Science Foundation of China (NSFC). A.S. is grateful to the Indian National Science Academy for his INSA Honorary Scientist position. R.B.H. was supported by Natural Environment Research Council (NERC) grant NE/V00249X/1 (Sat-Risk) and NERC National Public Good activity grant NE/R016445.

**Open Research**
**Data Availability Statement**
The IMP-5 proton fluxes for the August 1972 event are obtained from NASA's OMNI web (https://omniweb.gsfc.nasa.gov/). The Pioneer 10 IMF data for the August 1972 event are collected from NASA's COHO web (https://omniweb.gsfc.nasa.gov/coho/). The X-ray irradiances and proton fluxes for the March 1989 event are obtained from the GOES-06 satellite (http://www.ngdc.noaa.gov/stp/satellite/goes/dataaccess.html). Geomagnetic Dst, SYM-H, AE, and AL indices are collected from the World Data Center for Geomagnetism, Kyoto, Japan (https://wdc.kugi.kyoto-u.ac.jp/).